\documentclass[12pt]{iopart}
\usepackage{dcolumn}
\usepackage{bm}
\usepackage{latexsym}
\usepackage{amsfonts}
\usepackage[dvips]{graphicx}
\usepackage[usenames]{color}
\usepackage{makeidx}
\newcommand{\Slash}[1]{\ooalign{\hfil /\hfil\crcr$#1$}}

\newcommand{\pa}{\partial}

\newcommand{\ket}{\rangle }

\newcommand{\vphi}{\varphi}
\newcommand{\ve}{\varepsilon}

\newcommand{\up}{\uparrow}
\newcommand{\dw}{\downarrow}

\begin{document}

\title{Confinement Phase in Carbon-Nanotubes and the Extended Massive Schwinger Model
}

\author{Takashi Oka} 
\address{Max-Planck-Institut f{\"u}r Physik komplexer Systeme, N{\"o}thnitzer Stra{\ss}e 38, 01187 Dresden, Germany}
\address{Max-Planck-Institut f{\"u}r Chemische Physik fester Stoffe, N{\"o}thnitzer Stra{\ss}e 40, 01187 Dresden, Germany}
\address{The Institute for Solid State Physics, The University of Tokyo, Kashiwa, Chiba 277-8581, Japan}

\date{\today}
\begin{abstract}
\noindent 
Carbon nanotube with electric fluxes confined in
one dimension is studied. 
We show that a Coulomb interaction $\propto |x|$ 
leads to a confinement phase
with many properties similar to QCD in 4D. 
Low-energy physics is described by
the massive Schwinger model with multi-species fermions 
labeled by the band and valley indices.  
We propose two means to detect this state. 
One is through an optical measurement of the 
exciton spectrum, which has been calculated via the 
't Hooft-Berknoff equation 
with the light-front field theory. We show that 
the Gell-Mann$-$Oakes$-$Renner relation is satisfied by a
dark exciton.
The second is the nonlinear transport which 
is related to Coleman's ``half-asymptotic" state.
\end{abstract}

\maketitle

Even after two decades from its discovery, fascination continues with the 
carbon nanotube\cite{Iijima91}. 
One line of research is motivated by its
possible application as opto-electronic devices.
Another is more academic, in which 
possibilities are explored to realize 
new states of matter. 
In low dimensions 
quantum fluctuations are enhanced, which makes the 
nanotube an ideal host for 
strongly correlated and clean quantum systems. 
One monumental result along this line 
is the Tomonaga-Luttinger liquid 
in single-wall metallic nanotubes
\cite{Bockrath1999,EggerGogolin1997,KaneBalentsFisher1997}.
In the present work, we theoretically predict a
realization of yet another interesting state, 
namely the {\it confinement phase}, in nanotubes. 
In this state, charged 
particles cannot exist as free asymptotic states 
where all the excitations are neutral bound states 
(i.e., excitons in the present case).   

A well known realization of the 
confinement phase is the hadronic system 
where the effective quantum field theory 
is QCD in (3+1)D.  
By contrast, a  carbon nanotube is 
a one-dimensional object, and 
no dynamical nonabelian gauge field exists. 
Nonetheless, we propose here a confinement phase may exist;
A key is the dimensionality. 
In a seminal paper by Schwinger, 
a (1+1)D version of QED, 
{\it i.e.} the Schwinger model, 
was studied\cite{Schwinger62}.
In this model, Dirac particles interact through 
a one-dimensional Coulomb potential ($\propto |x|$).
The long-range interaction 
mimics the confinement potential, and
the groundstate is indeed in a confinement phase.  
The Schwinger model and its extensions were studied 
as a toy model of (3+1)D QCD to understand non-perturbative
aspects of the confinement phase \cite{CJS,Coleman:1976uz,Bergknoff77,Mo1993159,Harada1994,Harada1999,HetrickHosotaniIso}.
Recently, realization of the Schwinger model 
using ultracold atoms have been done
considered~\cite{Hauke2013PRX,Martinez2016Nature}. 
However, up to now, no solid-state realization of this model is known. 
Our claim here is that the low-energy effective
model for electrons in a nanotube can be, in certain 
situations, modeled by an 
extended massive Schwinger model. 
It is well known that a multi-species Dirac spectrum is realized
in the band structure of nanotube within the effective 
mass approximation. In order to realize the
one-dimensional Coulomb potential, the 
electric fluxes must be confined along the tube.  
The situation with shielded 1D electric flux can possibly be realized in 
(a) multi-wall nanotubes with metallic outer tubes, 
(b) a tube surrounded by metallic tubes, or 
(c) tubes embedded in a superconductor.  While 
one may wonder if such shielding can really take place, 
there is a prominent example in two-dimensional (2D) organic crystals
studied by Yamaguchi {\it et al.}\cite{Yamaguchi06}. 
They have experimentally shown that a layered 
structure with a large difference 
in the dielectric constants
leads to confinement of electric fluxes 
in the 2D plane where the Coulomb potential 
becomes logarithmic.  
In this experiment, the long-range Coulomb interaction leads 
to a power-law current-electric field ($J$-$E$) characteristics 
with a temperature dependent power. 
This gives us a strong motivation to study nanotube with a 
1D Coulomb potential. 
The properties of the confinement phase 
is reflected in the excitation spectrum. 
In order to explore this, we propose
two observable properties, one 
optical and another transport.

{\it Model ---}
We study nanotubes within the effective-mass formalism.  
There are infinitely many 
bands corresponding to different modes 
along the circumference of the tube, 
which we label with $n=0,\pm 1,\ldots$.  
The way in which the tube is wound is characterized by 
an index $\nu=0,\pm 1$,  which in turn specifies whether 
a discrete set of momenta  along the tube circumference intersect 
the two Dirac cones at $\alpha=$K, K' points in the graphene 
Brillouin zone ($\nu=0$; the ``(semi)metallic" 
case), or not 
($\nu=\pm 1$; the ``semiconducting"), for the half-filled band.   
In addition there is the spin degeneracy for $\sigma=\up, \dw$ 
with the Zeeman effect neglected here.  
Each band is characterized by 
a mass $\hbar v_F\kappa^\alpha(n)$, 
where $\kappa^{\rm K,K'}(n)=\frac{2\pi}{L}(n\pm \vphi-\nu/3)$
and $v_F$ the Fermi velocity in graphene\cite{AjikiAndo93}.
We take $\hbar v_F=1$ to be the unit of energy.
Here we have introduced a magnetic field 
whose flux passing through the tube is $\varphi$ 
in units of the flux quantum $\vphi_0=ch/e_0$, which 
acts to shift the discrete set of momenta. 
Assuming 1D electromagnetic fields,
the system is described by the extended massive Schwinger model 
with a Lagrangian
\begin{eqnarray}
\mathcal{L}=-\frac{1}{4}F_{\mu\nu}F^{\mu\nu}+
\sum_{n,\sigma,\alpha}
\bar{\psi}_{n,\sigma,\alpha}[i\Slash{\pa}-e\Slash{A}-\kappa^\alpha(n)]\psi_{n,\sigma,\alpha},
\label{eq:model}
\end{eqnarray}
where $F_{\mu\nu}=\pa_\mu A_\nu-\pa_\nu A_\mu$ is 
the electromagnetic field tensor, $\psi=(\psi_R,\psi_L)^T$ 
the fermion field with $\psi_R,\psi_L$ the left and
right moving components, $e=e_0/\sqrt{\ve_r}$ 
the screened charge with a dielectric constant $\ve_r$, 
and we use the relativistic convention
$g_{\mu\nu}=\mbox{diag}(1,-1)$ with 
$\gamma^0=(^{01}_{10}),\;\gamma^1=(^{0-1}_{1\; 0})$
and $\Slash{A}=A_\mu\gamma^\mu$. 
The model has a spin SU(2) symmetry, 
which is enlarged to SU(4) if K and K' are degenerate.  
However, perturbations
can break this into SU(2)$\times$SU(2).   
When the electric fluxes are confined in 1D, 
the inter-electron potential
is $V(x)=e^2|x|/2$, which enforces 
confinement of charges.  The coupling strength is 
$e^2/(\pi v_F)=0.709/\ve_r$ 
determined by the screening 
factor $\ve_r$ (see e.g. \cite{Andoexciton66}).

{\it Exciton spectrum ---}
Excitons are of central importance in 
understanding the optical properties of carbon-nanotubes
\cite{Ichida99,Connell02,Bachilo02,KishidaNanotube,WangNanotube}. 
Specifically, Kishida et al. have observed not only 
the bright excitons, but also the dark excitons\cite{KishidaNanotube}, while 
Wang et al. observed excitons in metallic nanotubes\cite{WangNanotube}. 
In the confinement phase, the optical spectrum is 
dominated by excitons, and no two-particle continuum exists. 
Since there is a possibility that the lightest fermion mass
vanishes, the conventional weak-coupling approaches cannot be used,
and strong-coupling methods are required.
Two powerful methods are applicable;
One is the light-front quantization scheme\cite{Bergknoff77,Mo1993159,Harada1994}, 
and the other is bosonization \cite{CJS,Coleman:1976uz,HetrickHosotaniIso}.
For studying the crossover from 
metallic to semiconducting tubes, we employ the former, 
since bosonization for a massive model is still an open problem\cite{HetrickHosotaniIso}.  
With the light-cone coordinates
$x^\mu=(x^+,x^-)\equiv
(x^0+x^1, x^0-x^1)/\sqrt{2}$ 
we can eliminate the dynamical variables except for 
the right-fermionic components
$\psi_{iR}$ (with a shorthand 
$i\equiv (n,\sigma,\alpha) $) by means of 
the equation of motion.
The Lagrangian reads
\begin{eqnarray}
&&L=\int dx^-\mathcal{L}=i\sqrt{2}\int dx^-\sum_i:\psi_{iR}^\dagger
\pa_+\psi_{iR}:\nonumber\\
&&+\frac{i}{2\sqrt{2}}\int dx^-dy^-\sum_i\kappa^2(n)\psi_{iR}^\dagger
(x^-)\ve(x^--y^-)\psi_{iR}(y^-)\nonumber\\
&&+\frac{e^2}{4}\int dx^-dy^-j^+(x^-)|x^--y^-|j^+(y^-)
\label{eq:lagrangian}
\end{eqnarray}
with the U(1) current
$j^\mu=\sum_i:\bar{\psi}_i\gamma^\mu\psi_i:,\; 
\gamma^+=(^{0\; 0}_{\sqrt{2}0}),\;\gamma^-=(^{0\sqrt{2}}_{0\; 0})$.
The index $i$ runs over infinite number of modes with the 
mass term depending on $n$, 
which contrasts with the standard SU($N$) massive Schwinger model
where the mass is common to all $i$'s.
The free-field expansion is 
$
\psi_{iR}(x^-)=\frac{1}{2^{1/4}}\int_0^\infty
\frac{dk^+}{2\pi\sqrt{k^+}}[b_i(k^+)e^{-ik^+x^-}
+d_i^\dagger (k^+)e^{ik^+x^-}],$
where $b^{\dagger}$ creating electrons and $d^{\dagger}$ 
holes satisfy a canonical commutation, 
$
\{b_i(k^+),b_j^\dagger (l^+)\}=\{
d_i(k^+),d_j^\dagger (l^+)\}=2\pi k^+\delta_{ij}\delta (
k^+-l^+).$
The virtue of using the light-front formalism is that 
the groundstate, which is a 
confinement phase (CP), is described by
the Fock vacuum $|0\ket_{\rm CP}=|0\ket_{\rm Fock}$ 
(with $b_j|0\ket_{\rm Fock}=d_j|0\ket_{\rm Fock}=0$). 

Now let us look at the two-particle 
excitation, i.e., an exciton with a wave function  
\begin{eqnarray}
|\psi\ket=\int_0^P\frac{dk_1dk_2}{2\pi\sqrt{k_1k_2}}
\sum_{i=1}\psi_i(k_1,k_2)b_{i}^\dagger (k_1)d_{i}^\dagger (k_2)|0\ket_{\rm CP},
\end{eqnarray}
where the integral is restricted to $k_1+k_2=P$.
From the Lorentz invariance, the exciton wave 
function satisfies the 
Einstein-Schr\"odinger equation, 
$2P^-P^+|\psi\ket=M^2|\psi\ket$, 
where $P^-$ is the light-cone
Hamiltonian, 
$P^+$ the momentum operator for 
the center of mass momentum $P$ with $P^+|\psi\ket=P|\psi\ket$, 
and $M$ the excitation energy (``mass") of the boundstate.
With the light-cone operators, the Einstein-Schr\"odinger equation 
for the wave function is given explicitly as 
\begin{eqnarray}
\frac{M^2}{2}\psi_i(\tilde{k},1-\tilde{k})=\left[\frac{\kappa^2(n)}{2}
-\frac{e^2}{2\pi}\right]\left(\frac{1}{\tilde{k}}+\frac{1}{1-\tilde{k}}\right)
\psi_i(\tilde{k},1-\tilde{k})\nonumber\\
-\frac{e^2}{2\pi}\int_0^1 d\tilde{k}'\frac{\psi_i(\tilde{k}',1-\tilde{k}')}{(\tilde{k}-\tilde{k}')^2}
+\frac{e^2}{2\pi}\int_0^1 d\tilde{k}'\sum_{j}\psi_j(\tilde{k}',1-\tilde{k}')
\label{eq:tb}
\end{eqnarray}
with re-scaled momenta $\tilde{k}=k/P,\;\tilde{k}'=k'/P$ \cite{derivation}. 
This is an extension of the 
't Hooft-Bergknoff equation\cite{'tHooft1974461,Bergknoff77}, 
here possessing infinite number of modes labeled by $i$.
The last term is the anomaly term, which physically corresponds to a
virtual process of exciton pair-annihilated into a photon
and then regenerated as an exciton. 
The process is intimately related to the photon-exciton coupling, 
and for bright excitons this
term is nonzero.  By contrast, the term 
disappears for dark excitons with 
$\sum_i\psi_i=0$, and eqn.(\ref{eq:tb}) reduces to the 
't Hooft equation for planar QCD\cite{'tHooft1974461}.
For the optical activity 
the bright excitons must satisfy the condition that 
$
\psi_{n\up\alpha}=\psi_{n\dw\alpha}
$
to be a SU(2) spin singlet (note $d^\dagger_{n\sigma l}$ creates
a hole with spin $-\sigma$), and, 
when K and K' are degenerate, an additional condition
$
\psi_{n\sigma {\rm K}}=\psi_{n\sigma {\rm K'}}
$
is imposed to make it a SU(2) 
valley singlet.  
We note that in eqn. (\ref{eq:tb}) the effect of 
vacuum polarization and self-energy corrections due to ``meson"
propagators \cite{Harada1999} are neglected for simplisity.
We solve\cite{commentbasisf} the 't Hooft-Bergknoff equation
using the basis-function method\cite{Mo1993159,Harada1994}.

\begin{figure}[tbh]
\centering 
\includegraphics[width=8.cm]{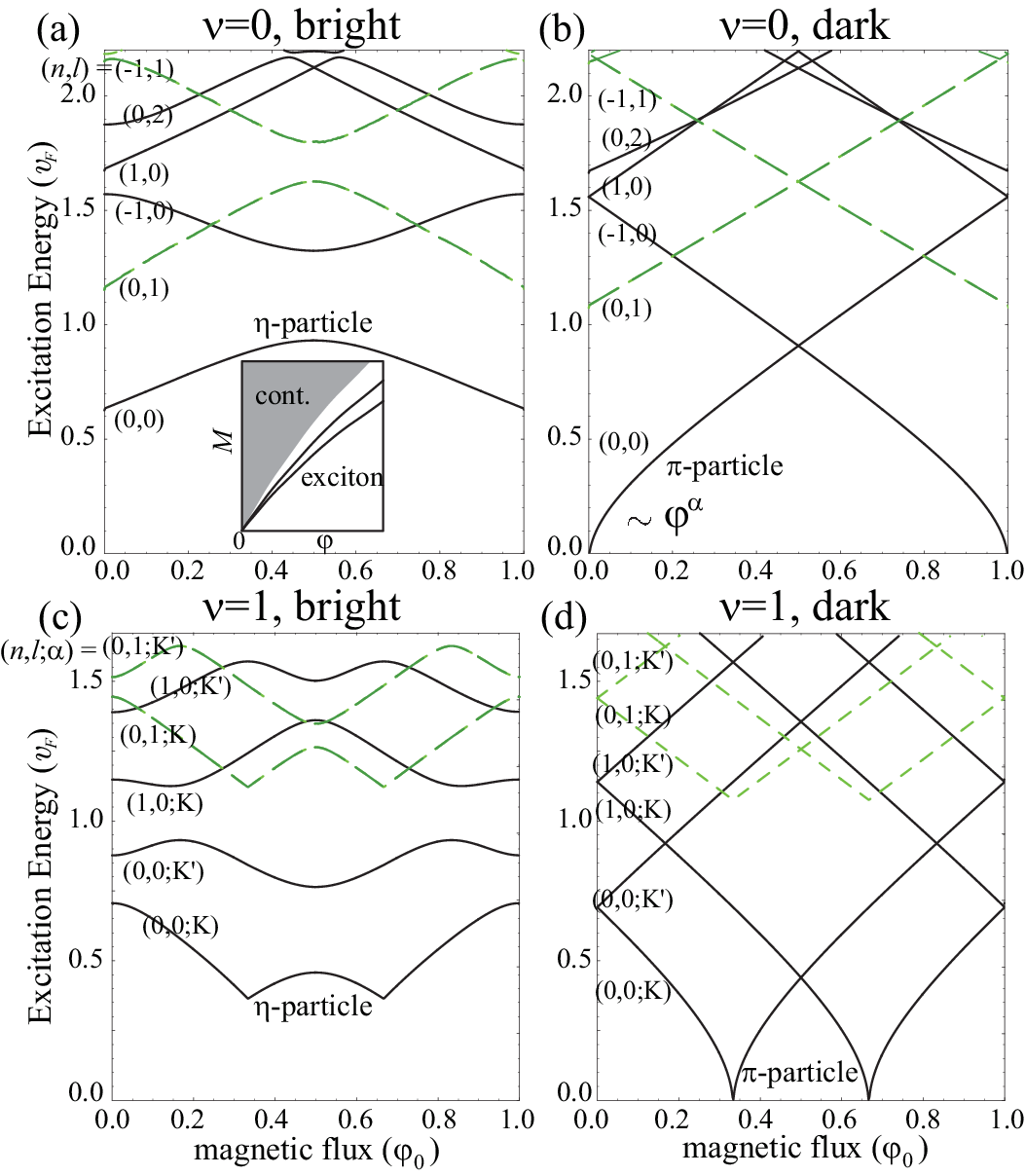}
\caption{(Color online)
Two-body excitation spectrum against magnetic field 
for bright (a,c) or dark (b,d) excitons for 
metallic (a,b; $\nu=0$) or semiconducting (c,d; $\nu=1$) 
nanotubes with $L=10,\;e^2/\pi v_F=0.2$.
Solid (dashed) lines correspond to 
states with even (odd) $l$.
Inset in (a) is a schematic spectrum 
for a system with $1/r$ potential near $\vphi=0$.
}
\label{fig:energyB}
\end{figure}
Let us first look at the excitation spectrum (exciton energy $M$) against the 
magnetic field for a metallic nanotube with $\nu=0$
in Fig.~\ref{fig:energyB}.  
We immediately notice that
the system is {\it no longer metallic due to 
charge confinement}, namely, the spectrum for the bright 
exciton has a gap, and an 
excitation continuum does not exist, either.  
This is in sharp contrast 
with the case for the conventional weak-coupling picture with 
a $1/r$ potential (inset of (a)),
where a continuum exists down to zero energy at $\vphi=0$.  
The spectrum has a periodicity with a 
period $\vphi=1$, 
which originates from the mass structure of the
fermion modes.
We label each exciton mode with 
$(n,l;\alpha)$, where 
 $\alpha=$K,K' is the dominant valley character near $\vphi= 0$,  
and $l=0,1,\ldots$  the exciton quantum number
that labels the bound state in a trapping potential (see Fig.~\ref{fig3}~(c)).  
Odd-$l$ states are parity odd and one-photon allowed, while
even-$l$ states are only two-photon accessible.
For $\nu=0$ the $(n,l;K)$
and $(-n,l;K')$ excitons are degenerate due to the 
valley symmetry, 
so we can omit $\alpha$ from the index.

\begin{figure}[tbh]
\centering 
\includegraphics[width=8.cm]{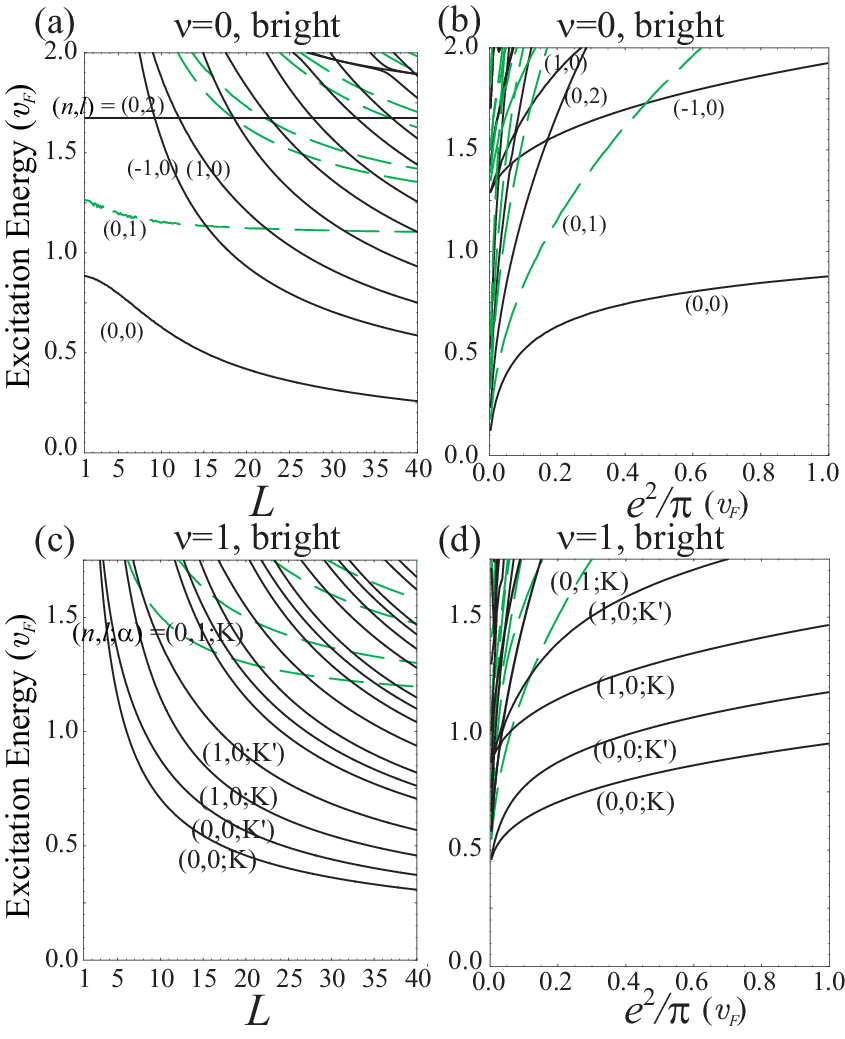}
\caption{(Color online)
Two-body excitation spectrum in zero magnetic 
field against circumference $L$ (a,c; 
with $e^2/\pi v_F=0.2$) or 
against coupling strength (b,d; with $L=10$) 
for metallic (a,b; $\nu=0$) or  
semiconducting (c,d; $\nu=1$) nanotubes.
Solid (dashed) lines correspond to 
states with even (odd) $l$.
}
\label{fig2}
\end{figure}

What can we learn more about the spectrum?
Comparison with the meson spectrum in QCD becomes interesting. 
In fact, the bright and dark excitons with 
lowest energies (Fig.\ref{fig:energyB}) have many aspects shared by the 
$\eta$ and $\pi$ (pion) particles, 
respectively, in QCD\cite{Coleman:1976uz}. 
Interestingly, the $\eta$ and $\pi$  boundstates behave
differently in the strong-coupling limit;
As the fermion mass goes to zero,
the $\pi$ mass vanishes, while the $\eta$
mass remains finite due to a U(1) anomaly. 
For QCD there is the 
{\it Gell-Mann$-$Oakes$-$Renner relation}\cite{GOR}, 
which is a relation, 
$M_{\pi} \propto m^{1/2}_{\rm quark}$, 
between the pion and quark masses.
When translated into the present problem, the relation 
applies to the lightest dark exciton. 
For example, as shown 
in Fig.~\ref{fig:energyB} (b,d), 
the energy of the lightest dark exciton, i.e.,
pion in the QCD context, 
goes to zero at $\vphi=\bar{\vphi}=0,1$ in metallic nanotubes 
and at $\bar{\vphi}=1/3,\; 2/3$ in semiconducting ones. 
However, the exciton energy behaves as  
$M\propto |\vphi-\bar{\vphi}|^{\alpha}$
with a power $\alpha\simeq 0.5$ and, since the 
fermion mass is proportional to $|\vphi-\bar{\vphi}|$,
we can regard this as a manifestation of the 
{\it Gell-Mann$-$Oakes$-$Renner relation in nanotubes},
which holds even in 1+1D systems 
where chiral symmetry is `almost' broken \cite{Witten1978}.

The situation is even more interesting for the lightest
bright excitons ($\sim \eta$-particles), since they remain 
massive even though the fermions are massless. 
This can be seen in the $(0,0)$ and $(0,0;K)$ states 
in Fig.~\ref{fig:energyB} (a)~(c), 
which have nonzero minima at $\vphi=\bar{\vphi}$.
This is due to a U(1) {\it anomaly} coming from the pair creation-annihilation process, 
i.e., the last term in eqn.~(\ref{eq:tb}). 
The physical picture is the following. 
In an exciton, electrons and holes are continuously created and annihilated
when the fermion mass is small, and a cloud of photon is formed. 
The electromagnetic energy of the
photon cloud is finite, and contributes to the 
exciton energy. 
This process is not restricted to the 
1D Coulomb potential, and similar effect may take place 
in metallic single-wall nanotubes where
excitons with finite binding energy were observed \cite{WangNanotube}.

In experiments it often happens that the nanotubes have random circumference $L$. 
In Fig.\ref{fig2}(a,c), we plot the dependence of the 
exciton energy on the circumference.
The mixing between different modes $n$ 
is small when $L$ is small, 
since the energy difference between modes is $\propto 1/L$, and 
in the limit $L\to 0$, the system 
approaches to the pure SU($N$) massless Schwinger model \cite{Coleman:1976uz}.
In this limit, the lightest bright exciton 
($\eta$-particle) mass is given by $M_\eta=\sqrt{Ne^2/\pi}$,
which amounts to $M_\eta= 0.8944$ 
in metallic ($\nu=0$, SU(4)) nanotubes 
and $M_\eta= 0.6325$ in 
semiconducting  ($\nu=1$, SU(2)) ones 
for $e^2/\pi v_F=0.2$.  
The mixing of modes lowers the 
exciton energy as shown in Fig.\ref{fig2} (a,c).
This is because the mixing reduces the 
anomaly term, where the $(0,0)$ mode converges to the
dark mode in the SU($\infty$) 
massless Schwinger model.
In Fig.\ref{fig2} (b,d), we plot the 
dependence of the spectrum 
on the screened interaction parameter $e^2$. 
Starting from $2\kappa$ (weak-coupling limit),
the exciton energy increases as the 
interaction becomes stronger, 
where the increase is larger for 
larger quantum number $l$.

\begin{figure}[hbt]
\centering 
\includegraphics[width=8.5cm]{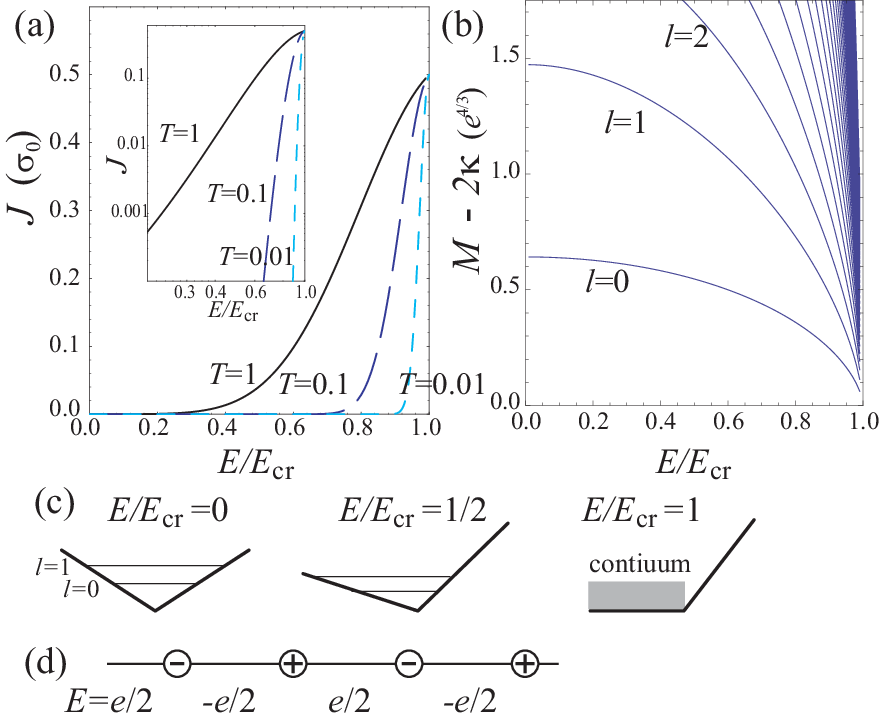}
\caption{(Color online)
$J$-$E$ characteristics(a) with a logarithmic plot in the inset, 
and exciton levels(b) 
for semiconducting nanotubes (for 
$\xi=100$)
in the weak-coupling, nonrelativistic limit.  
(c) Schematic trapping potential $V(x)$
and eigenstates for various values of the electric field.
(d) Coleman's half asymptotic state when the external 
field is $E=E_{\rm cr}$.
}
\label{fig3}
\end{figure}

{\it Nonlinear transport and half-asymptotic state ---}
Nonlinear transport in correlated systems is attracting 
considerable interests (see e.g. \cite{okaprl}), 
and in nanotubes it serves as a powerful probe in experimentally 
detecting new states of matter \cite{Bockrath1999}.
Here let us show that 
the semiconducting nanotubes with 
1D interaction should exhibit a 
power-law $J$-$E$ characteristics in external electric fields, 
as in the 2D organic systems\cite{Yamaguchi06}. 
Unlike the 2D case, however, 
the carriers in the field-induced metallic state
are only ``half-free".  We explain how 
this has to do with the half-asymptotic state
predicted by Coleman\cite{Coleman:1976uz}.

Let us look at a 
semiconducting nanotube with a finite fermion
mass $\kappa$ in a strong electric field. 
Since the strong-coupling analysis
in finite electric fields is still an open issue,
here we focus on a weak-coupling approach
where the U(1) anomaly does not contribute 
with the bright and dark excitons 
becoming nearly degenerate.
In the nonrelativistic approximation, 
Dirac particles becomes nonrelativistic fermions
with quadratic kinetic terms ($p^2/2M$),
and the exciton binding problem in an electric field
$E$ is reduced to solving a 
1D Schr\"odinger equation (c.f. eqn. (4.1) in \cite{Coleman:1976uz}), 
\begin{eqnarray}
&&[-\pa_x^2+V(x)]\psi^{(l)}(x)=(M_l-2\kappa)\psi^{(l)}(x),
\label{eq:schrodinger}
\\
&&
V(x)=\frac{e^2}{2}|x|e^{-|x|/\xi}+e_0Ex,
\end{eqnarray}
where $M_l$ is the exciton energy, 
$V(x)$ the trapping potential,
and we have introduced an exponential damping factor with 
a cutoff $\xi$.  
We plot the $J$-$E$ characteristics in Fig.~\ref{fig3}(a)
for various values of temperature, 
where the nonlinear current is seen to behave like 
\begin{eqnarray}
J=\sigma_0\exp[-\Delta(E)/(2k_BT)]E,
\label{eq:current}
\end{eqnarray}
as in ref.\cite{Yamaguchi06}
with $\Delta(E)$ the activation energy.
The current follows a power-law 
until the conductivity reaches a peak 
at a critical field $E=E_{\rm cr}\equiv e_0/2\ve_r$. 
The power of the conductivity is proportional
to $T^{-1}$ as in ref.~\cite{Yamaguchi06} but there is also a 
cutoff $\xi$ dependence.
At the critical field, one side of the trapping potential
for a test charge becomes flat 
as depicted in Fig.~\ref{fig3}(c). 
Then a continuum spectrum emerges, 
as seen in Fig.~\ref{fig3}(b) 
where we plot the eigenenergies of the 
Hamiltonian (eqn.(\ref{eq:schrodinger})) 
obtained by gluing two Airy functions. 
There, Coleman's half-asymptotic state --
a configuration with alternating charges 
but possibly random displacements (Fig.~\ref{fig3}(d))
-- 
has the lowest energy, where the external field and the
force from surrounding charges balance with each other. 
Strictly speaking, carriers with
opposite charges must switch their
position in order for the current to 
flow, and this violates
the half-asymptotic state condition, 
which may modify the simple 
relation eqn.(\ref{eq:current})
for the nonlinear current.   
This is out of the scope of the present 
work, but will merit further studies.

TO wishes to thank Kazuhiro Kuroki 
for valuable discussions, and 
was supported by JST CREST Grant No. JPMJCR19T3, Japan.


\begin{thebibliography}{99}
\bibitem{commentbasisf} 
We expand the wave function as
$\psi_i(\tilde{k},1-\tilde{k})=\sum_{j=0}^{N_{\rm cut}}
a_{nj}f_j(\beta_n,\tilde{k})$,
where $f_j(\beta,\tilde{k})$ is $[\tilde{k}(1-\tilde{k})]^{\beta+j}$
for parity even and
$[\tilde{k}(1-\tilde{k})]^{\beta+j}(2\tilde{k}-1)$
for odd solutions.
$\beta_n$ is a solution for $\frac{\pi}{e^2}\kappa^2(n)-1
+\pi\beta_n\cot(\pi\beta_n)=0$, 
which is here $n$ dependent.
\end{thebibliography}

\end{document}